\documentstyle[aps,preprint]{revtex}

\newcommand{\be}{\begin{equation}}
\newcommand{\ee}{\end{equation}}
\newcommand{\ba}{\begin{eqnarray}}
\newcommand{\ea}{\end{eqnarray}}
\newcommand{\ket}[1]{| {#1} \rangle}
\newcommand{\bra}[1]{\langle {#1} |}

\newcommand{\opoh}{{\it 1p1h }}
\newcommand{\tpth}{{\it 2p2h }}
\newcommand{\npnh}{{\it npnh }}
\newcommand{\GDR}{{\it GDR }}
\newcommand{\DGDR}{{\it DGDR }}
\def\lsim{\mathrel{\rlap{\lower4pt\hbox{\hskip1pt$\sim$}}
    \raise1pt\hbox{$<$}}}     
\def\gsim{\mathrel{\rlap{\lower4pt\hbox{\hskip1pt$\sim$}}
    \raise1pt\hbox{$>$}}}     

\begin{document}

\draft

\title{DOUBLE-DIPOLE EXCITATIONS IN $^{40}$Ca}

\author{S. Nishizaki\cite{byline1} and J. Wambach\cite{byline2}}

\address{Department of Physics\\
University of Illinois at Urbana-Champaign\\
1110 W. Green St., Urbana, IL 61801}

\maketitle

\begin{abstract}
The double-dipole strength distribution in $^{40}$Ca is
calculated microscopically within a model space
of \opoh - and \tpth excitations. Anharmonic effects
in the centroid energies of the $0^+$- and $2^+$ components
are found to be small, in agreement with
experimental observation. Firm conclusions about the
spreading width cannot be drawn, as yet, due to computational
limitations in the number of \tpth states.
\end{abstract}

\section{Introduction}
In pion double-charge-exchange reactions as well as in
peripheral heavy-ion collisions at relativistic bombarding energies
$\gsim$ 1 GeV/nucleon the double-giant-dipole resonance ({\it DGDR})
is strongly excited (see~\cite{Mord,Emli} for recent reviews). In the
latter case there
is a strong longitudinal focusing of the electromagnetic field in the
target rest frame. This greatly enhances the field intensity in the
vicinity of the target nucleus thus increasing the probability for
two-photon absorption from the ground state. The situation is somewhat
similar to multi-photon excitations of atoms with intense laser pulses.

Most simply, the \DGDR is understood as a two-phonon excitation
(with $J^{\pi}=0^+,2^+$ in a spherical nucleus) which, in the harmonic
limit, is located at twice the giant dipole resonance ({\it GDR}) energy.
This expectation is largely confirmed by experiment~\cite{Emli}.
On the other hand, the single-dipole state
\be
\ket{D}\equiv\sum_{i=1}^Z\vec{r}_i\tau^z_i\ket{0},
\ee
as well as the double-dipole state
\be
\ket{DD}\equiv\bigl(\sum_{i=1}^Z\vec{r}_i\tau^z_i\bigr)
\bigl(\sum_{j=1}^Z\vec{r}_j\tau^z_j\bigr)\ket{0},
\ee
are not eigenstates of the nuclear Hamiltonian and hence acquire a width.
Here the experimental situation is less clear and values for $\Gamma_{DGDR}$
are observed which are bracketed by $\sqrt{2}\Gamma_{GDR}$ and
$2\Gamma_{GDR}$ \cite{Emli}. The lower value is expected from a folding
of two Gaussians while the larger width results from the folding of two
Lorentzians (recall that the Lorentzian gives a good fit to the
total photoabsorption cross section).

To reach a quantitative understanding of the \DGDR characteristics a
microscopic description is needed. A first calculation
within the second RPA formalism \cite{JaDG,DNSW} was presented in
ref.~\cite{LaRe} using a separable residual interaction
and various degrees of approximation. More recently~\cite{Pono},
particle-vibration coupling calculations of the semiclassical
Coulomb excitation cross section
have been reported which, by construction, select a restricted set of
\tpth diagrams and have difficulties with properly imposing the Pauli
principle. In the present paper we wish to present a calculation
of the single- and double-dipole response function which
diagonalizes a given model Hamiltonian, $\hat H$, in the space of
\opoh - and \tpth excitations avoiding diagrammatic selections and fully
respecting the Pauli principle. The theory has been used previously
to provide a realistic description of linear response functions over a
wide range of excitation energies \cite{DNSW}.

\section{Theoretical Development}

We start by expanding the wave packets $\ket{D}$ and $\ket{DD}$
(eqs.~(1) and (2)) in exact eigenstates $\ket{N}$ of the nuclear Hamiltonian
$\hat H$ as
\ba
\ket{D}=\sum_{N\ne 0}\bra{N}\hat D\ket{0}\ket{N}\nonumber\\
\ket{DD}=\sum_{N\ne 0}\bra{N}\hat D\hat D\ket{0}\ket{N}
\ea
where $\hat D=\sum_{i=1}^Z\vec{r}_i\tau^z_i$.
When writing $\hat H$ as a mean field part and a residual interaction
\be
\hat H=\sum_i\epsilon_i a_i^{\dag} a_i+{1\over 4}\sum_{ij,kl}v_{ij,kl}
a_i^{\dag} a_j^{\dag} a_la_k.
\ee
($v_{ij,kl}$ denote antisymmetrized two-body matrix elements)
the exact eigenstates $\ket{N}$ are obtained by diagonalization in the
space of \npnh excitations. A minimal truncation of this,
prohibitively large, space which can accommodate both single- and double dipole
excitations must clearly encompass the \opoh and \tpth sector, so
that
\be
\ket{N}=\sum_{1}c^N_1\ket{1}+\sum_2c^N_2\ket{2} .
\ee
Due to the one-body nature of the dipole operator, the overlap matrix elements
$\bra{N}\hat D\ket{0}$ then project onto the \opoh subspace, $\ket{1}$, with
\be
\bra{N}\hat D\ket{0}=\sum_1c^{N*}_1\bra{1}\hat D\ket{0}
\ee
while $\bra{N}\hat D\hat D\ket{0}$ project onto the \tpth space, $\ket{2}$,
with
\be
\bra{N}\hat D\hat D\ket{0}=\sum_2c^{N*}_2\bra{2}\hat D\hat D\ket{0}.
\ee
Knowing the expansion coefficients, $c^N_{1,2}$ and the eigenvalues $E_N$
of the Hamiltonian matrix in these spaces one then readily computes
the `strength functions`
\ba
S_D(E)=\sum_N|\bra{N}\hat D\ket{0}|^2\delta(E-E_N)\nonumber\\
S_{DD}(E)=\sum_N|\bra{N}\hat D\hat D\ket{0}|^2\delta(E-E_N).
\ea
The angular-momentum-coupled expression for $S_{DD}$ is somewhat involved
and will be given elsewhere \cite{NiW1}. For comparison with experiment it is
useful to define the centroids and variances of the strength distributions
(8) in terms of their energy moments
\be
m^k_{D(DD)}=\int dE E^kS_{D(DD)}(E)
\ee
as
\be
E_{GDR}=m^1_D/m^0_D;\quad \sigma_{GDR}=\sqrt{m^2_D-(m^1_D)^2}/m^0_D
\ee
and similarly for the double-dipole excitation.

\section{Results}

Although not measured, so far, we have chosen the $^{40}$Ca nucleus since
the dimensions of the model space can be kept under reasonable
numerical control. For the mean field we have chosen the same Woods-Saxon
potential as in ref.~\cite{SW}. Continuum excitations have been treated by
discretization through an expansion of the positive-energy single-particle
states in a harmonic oscillator basis. The density-dependent zero-range
interaction of ref.~\cite{SW} has also been adopted. Its form ensures the
Pauli principle. To get a quantitative description of the \GDR we have
increased the isovector part, to 300 MeV fm$^3$ for, both, the interior
and exterior part of $V_{01}$. When truncating to \opoh
and \tpth configurations with excitation energies $\leq$ 30 MeV
(which yields 375 states) one obtains the
strength distribution displayed in the upper part of Fig.~1. The centroid
energy, $E_{GDR}$ is at 19.7 MeV as compared to 20.3 MeV, deduced the
global A-dependence $E_{GDR}=31.2 A^{-1/3}+20.6 A^{-1/6}$ \cite{BeFu}.
Upon further increase of the
model space, both the total transition strength and the centroid energy
remain unchanged. For further comparison, the lower part of Fig.~1 shows
the normalized photo absorption cross section \cite{RiSc}
\be
\sigma_\gamma(E)=\sum_N|\bra{N}\hat D\ket{0}|^2E_N\delta(E-E_N)/m^1_D
\ee
together with data from Compton scattering \cite{WDMN} (for the
theoretical distribution a Lorentzian smoothing of 1 MeV has been used).
While not reproduced in all detail the overall agreement with experiment
is quite satisfactory.

In the harmonic limit, the double-dipole state is expected at
$2\times E_{GDR}$ and a truncation at 30 MeV excitation energy is clearly
insufficient. On the other hand the density of \tpth states increases
rapidly with excitation energy. A reasonable compromise is to truncate
at 45 MeV (where the number of states for the $0^+$-component of the
DGDR is 1067 while for the $2^+$-component one obtains 4144 states). At
45 MeV the total transition strength as well as the centroid energy are
saturated. Any further increase in truncation energy will thus
only influence the variance, $\sigma_{DGDR}$, {\it i.e.} the width.
To test harmonicity one can compare the \DGDR centroid energies
$E^{0^+}_{DGDR} =38.67$ MeV and $E^{2^+}_{DGDR}=38.38$ MeV to
$2\times E_{GDR}=39.34$ MeV, indicating that the anharmonicity is quite
small. This is corroborated by the fact that the $0^+$- and $2^+$ components
are nearly degenerate. The normalized \DGDR strength distributions,
(Fig.~2), indicate some differences between the $0^+$- and
$2^+$ components, however, in fine structure as well as in the overall
width (the $0^+$-component is not strongly excited in experiment \cite{Eml1}).

To decide whether $\Gamma_{DGDR}/\Gamma_{GDR}$ is closer to
$\sqrt{2}$ or 2 we have determined the ratio of variances
$\sigma_{GDR}/\sigma_{DGDR}$ yielding 1.61 for the $0^+$-component and 1.24
for $2^+$ component ($\sigma_{GDR}$ was evaluated in a window from 10-30 MeV
while $\sigma_{DGDR}$ was obtained in a window from 30-50 MeV). It would be
premature to conclude that a ratio of $\sqrt{2}$ is favored, however, since
the model space might still be too small to get a reliable answer of the
width. We estimate that this limitation can be overcome. Before
embarking on large-scale calculations, however, one should evaluate the Coulomb
excitation cross section for direct comparisons with experiment \cite{NiW2}.

\section{Summary}

To get a quantitative assessment of the \DGDR, which has been observed
in several nuclei \cite{Emli}, we have presented a microscopic calculation
of the transition strength distribution in $^{40}$Ca within the space of
\opoh- and \tpth excitations. By comparing the mean energies of the single-
and double-dipole response it is found that anharmonicity effects are quite
small, in good agreement with experiment. For the respective widths the
conclusions are still hampered by computational restrictions of model
space. Before enlarging the space it would be desirable to
have a proper description of the Coulomb excitation cross section
\cite{Pono}. As another improvement, RPA ground-state correlations should be
included. While this is trivial in the \opoh sector, the \tpth sector is more
challenging numerically \cite{NiW2}.

\noindent

\bigskip

\noindent
{\bf Acknowledgement}:
We thank G. Baur, P. Debevec and H. Emling for fruitful discussions.
We also appreciate help from A. Trellakis in some of the numerical
work. This work is supported in part by the National
Science Foundation under Grant No. NSF PHY89-21025.

\newpage

\begin{center}
{\large Figure Captions}
\end{center}
\begin{itemize}
\item[Fig.~1:]
upper part: the calculated dipole transition strength distribution
in $^{40}$Ca (eq.~(8)). The high-energy tail has been multiplied by a factor
of 10; lower part:  the photoabsorption cross section,
normalized to the TRK sum rule. The data, indicated by the filled squares,
are inferred from the Compton-scattering analysis of ref.~[9].

\item[Fig.~2:]
The transition-strength distribution for the $0^+$-component
(dashed line) and $2^+$-component (full line) the \DGDR in $^{40}$Ca.
For comparison the single-dipole strength function is also given
short-dashed line). In all cases the discrete spectra have been folded
with a 1 MeV-width Lorentzian (the experimental resolution is typically
1-2 MeV [1].
\end{itemize}

\end{document}